# Gapless quantum spin liquid ground state in the two-dimensional spin-1/2 triangular antiferromagnet YbMgGaO$_4$

Yuesheng Li[1], Haijun Liao[2], Zhen Zhang[3], Shiyan Li[3], Feng Jin[1], Langsheng Ling[4], Lei Zhang[4], Youming Zou[4], Li Pi[4], Zhaorong Yang[5], Junfeng Wang[6], Zhonghua Wu[7], and Qingming Zhang[*,1,8]

[1]Department of Physics, Renmin University of China, Beijing 100872, P. R. China

[2]Institute of Physics, Chinese Academy of Sciences, Beijing100190, P. R. China

[3]State Key Laboratory of Surface Physics, Department of Physics, and Laboratory of Advanced Materials, Fudan University, Shanghai 200433, P. R. China

[4]High Magnetic Field Laboratory, Chinese Academy of Sciences, Hefei 230031, P. R. China

[5]Institute of Solid State Physics, Chinese Academy of Sciences, Hefei 230031, P. R. China

[6]Wuhan National High Magnetic Field Center, Wuhan 430074, P. R. China

[7]Institute of High Energy Physics, Chinese Academy of Science, Beijing 100049, P. R. China

[8]Department of Physics and Astronomy, Collaborative Innovation Center of Advanced Microstructures, Shanghai Jiao Tong University, Shanghai 200240, P. R. China

[*]e-mail: qmzhang@ruc.edu.cn

**Quantum spin liquid (QSL) is a novel state of matter which refuses the conventional spin freezing even at 0 K. Experimentally searching for the structurally perfect candidates is a big challenge in condensed matter physics. Here we report the successful synthesis of a new spin-1/2 triangular antiferromagnet YbMgGaO$_4$ with *R$\bar{3}$m* symmetry. The compound with an ideal two-dimensional and spatial isotropic magnetic triangular-lattice has no site-mixing magnetic defects and no antisymmetric Dzyaloshinsky-Moriya (DM) interactions. No spin freezing down to 60 mK (despite θ$_w$ ~ -4 K), the low-*T* power-law temperature dependence of heat capacity and nonzero susceptibility suggest that YbMgGaO$_4$ is a promising gapless (≤ |θ$_w$|/100) QSL candidate. The residual spin entropy, which is accurately determined with a non-magnetic reference LuMgGaO$_4$, approaches zero (< 0.6 %). This indicates that the possible QSL ground state (GS) of the frustrated spin system has been experimentally achieved at the lowest measurement temperatures.**

Low-spin geometrically frustrated systems in two-dimensional (2D) lattices have received significant interest in condensed-matter physics. The two most studied frustrated spin systems are spin-1/2 triangular and kagomé antiferromagnets, in which strong quantum fluctuations prevent spin freezing even at very low temperatures[1]. With respect to theoretical studies, Anderson first proposed that the triangular Heisenberg antiferromagnet (THAF) has a resonating valence bond GS, which is a type of spin liquid[2,3]. However, recent numerical studies have consistently indicated a long-range Néel GS for spin-1/2 THAF. The calculated order parameter is much smaller than the classical value, indicating that it is very close to a quantum critical point between magnetic ordered and disordered GSs[4,5]. On the experimental side, "structurally perfect" triangular or kagomé antiferromagnets (AFs) are still extremely rare, although many spin-1/2 geometrically frustrated triangular and kagomé AFs have been proposed[6-12]. Most of the existing candidates suffer from spatially anisotropic intralayer exchange interactions[6,7,9,12], site-mixing between magnetic and nonmagnetic ions[12-14], interlayer exchange interactions[8-10], and/or antisymmetric Dzyaloshinsky-Moriya (DM) interactions[15]. These factors are critical to the GSs of the frustrated spin systems and are difficult to determine precisely. The complications caused by these factors make extraction of the intrinsic physics from real systems difficult.

In this paper, we report the successful synthesis of a new triangular antiferromagnet with effective spin-1/2, YbMgGaO$_4$. The aforementioned structural disadvantages are avoided in the new compound. First, it has spatially isotropic and perfect triangular layers with $R\bar{3}m$ symmetry. Second, the number of magnetic defects is negligible because of the large chemical difference between the Kramers magnetic Yb$^{3+}$ ions and the nonmagnetic ions. Third, the magnetic triangular layers are well magnetically separated by nonmagnetic double layers of Mg/GaO$_5$ triangular bipyramids, indicating an ideal two-dimensionality of the magnetic layers and a negligible interlayer exchange interaction. Finally, the antisymmetric DM interactions between first-, second- and third-neighbor spins are strictly excluded because inversion centers are located at any $Yb^{3+}$ ion and at the half-way sites between them[16]. Moreover, the nonmagnetic reference compound $LuMgGaO_4$ is also available for control experiments, such as precisely excluding the lattice heat capacities for $YbMgGaO_4$.

No magnetic ordering is observed at least down to 60 mK from both magnetization and heat capacity measurements, despite the obvious AF exchange interaction between nearest-neighbor spins ($\theta_w \sim$ -4 K) suggesting that $YbMgGaO_4$ is a new QSL candidate. And the power-law temperature dependence of heat capacity and nonzero susceptibility further indicate that the excitation gap from the GS should be no more than $\sim |\theta_w|/100$ at low temperatures[17,18]. Almost zero residual spin entropies are observed under 0 to 9 T at $T$ < 0.3 K. This experimentally indicates that the frustrated spin system extremely approaches a possible gapless QSL GS at low temperatures. Moreover, an anomalous susceptibility plateau is first observed at paramagnetic states (0.5 K), which imply an unusual field-induced quantum spin state.

$YbMgGaO_4$ and $LuMgGaO_4$ are members of the $Ln^{3+}M^{2+}M'^{3+}O_4$ family ($R\bar{3}m$ symmetry, $a \sim$ 3.4 Å and $c \sim$ 25 Å), where Ln is Lu or Yb, and M and M' are 3d transition metals. Spin frustration has been observed in the other members of this structural family, including $YbCuGaO_4$, $LuCuGaO_4$, $LuCoGaO_4$, $LuZnFeO_4$ and $LuCuFeO_4$[19]. Unfortunately, the chemical disorder in these compounds, along with the geometrical frustration, resulted in spin-glass behavior[19]. Compounds with magnetic ions occupying only the lanthanide sub-lattice (triangular layer), such as YbZnGaO$_4$, have not been reported thus far. Our experiments demonstrate that pure YbZnGaO$_4$ is difficult to synthesize in air because ZnO is volatile at $T > 1200$ °C. However, MgO remains stable at $T < 1500$ °C, as confirmed by the almost zero mass loss of this compound after heating. This thermal stability of MgO allows us to successfully synthesize pure YbMgGaO$_4$.

For accuracy, we employed two starting crystal structure models: YbFe$_2$O$_4$ (I) and LuCuGaO$_4$ or YbCoGaO$_4$ (II)[19]. In model I, Ln$^{3+}$ locates at its ideal position (0, 0, 0). We fit the observed XRD intensities with reasonable refined parameters and small residuals in the Supplementary Information. In the case of the alternative model II, Ln$^{3+}$ slightly deviates from its ideal position to (0, 0, $z_{Ln}$), where $z_{Ln} \sim 0.004$. In this case, we also achieved a reasonable refinement using model II. However, an extra refinement parameter $z_{Ln}$ was required. Even when the deviation exists, the maximum displacement along the $c$-axis is only approximately $\pm 0.004c$ (0.1 Å). This possible displacement may be due to Mg$^{2+}$-Ga$^{3+}$ disorder in the interleaved double layers (Fig. 1a) and may cause slight bond randomness. Other researchers have reported that $z_{Ln}$ is $\sim 0.009$ in LnCuGaO$_4$, whereas $z_{Lu}$ is $\sim 0.006$ in LuCoGaO$_4$ and $z_{Yb}$ is $\sim 0.005$ in YbCoGaO$_4$[19,20]. These results are consistent with the fact that Cu$^{2+}$ is Jahn-Teller active, whereas Mg$^{2+}$, Co$^{2+}$ and Ga$^{3+}$ are not. Notably, the strict $R\bar{3}m$ symmetry is maintained in both cases; i. e., a triangular spin lattice remains spatially isotropic.

The crystal structure is shown in Fig. 1a. Triangular layers of magnetic YbO$_6$ octahedra, interleaved with double layers of nonmagnetic Mg/GaO$_5$ triangular bipyramids, are ABC-stacked along the *c*-axis. Thus, the interlayer magnetic coupling can be negligible compared to the intralayer superexchange interaction between neighboring $Yb^{3+}$, which is directly mediated by two parallel anions, $O1^{2-}$. Site-mixing between magnetic and nonmagnetic ions is forbidden because the radius of $Yb^{3+}$ is much larger than that of $Mg^{2+}$ or $Ga^{3+}$. This result is confirmed by the lack of observable narrow electron spin resonance (ESR) signals of isolated $Yb^{3+}$ ions in $YbMgGaO_4$ (Supplementary Information). The spatially isotropic triangular layer of $YbO_6$ is shown in Fig. 1b. Because the inversion centers are located at halfway sites between neighboring $Yb^{3+}$ ions (Fig. 1b) and $Yb^{3+}$ sites (Fig. 1c), the DM interactions between the first-, second-, and third- neighbor $Yb^{3+}$ vanish according to Moriya's rules[16]. Since $Yb^{3+}$ is a Kramers ion, the low-T ($\leq$ 30 K, see below) magnetism of $YbMgGaO_4$ should be dominated by the Kramers doublet GSs with effective spin-1/2.

The magnetizations are strongly suppressed with increasing $Yb^{3+}$ concentration or effective exchange coupling (Fig. 2a, b); this result suggests an AF neighbor exchange energy. The AF coupling is confirmed by the negative Weiss temperatures, $\theta_w < 0$ (Fig. 2c). The Curie-Weiss fit gives the Weiss temperature, $\theta_w = -4.11(2)$ K, and the averaged g-factor, 3.2(1), for $YbMgGaO_4$. Saturation behavior is observed in the MH curves below 10 T (Fig. 2b), whereas a linear dependence emerges above 10 T; this behavior is attributed to Van Vleck magnetism[8] and is related to field-induced electronic transitions. The saturation magnetizations are in agreement with the above averaged g-factor, $M_s = 1.600(2)$ $\mu_B/Yb^{3+}$ ~ $g_{ave}/2$, where the powder-averaged g-factor, $g_{ave}$ ~ 2/3 $g_\perp$ +1/3 $g_{//}$. In addition, a Van Vleck susceptibility (the fitted slope), $\chi_{vv}$ = 0.0122(1) $\mu_B$/Yb$^{3+}$/T, is also obtained.

Experimentally determining the GS of YbMgGaO$_4$ is of particular interest and non-trivial. The susceptibilities measured under zero field cooling (ZFC) and FC show no observable differences at least down to 0.48 K (Fig. 2d). A complete magnetization loop (-7 T to 7 T) was also measured at 0.5 K, and no hysteresis was observed (inset of Fig. 2d). The heat capacities measured under 0 to 9 T show no sharp λ-type peak down to 60 mK (Fig. 3a, b). These observations are incompatible with the long-range AF/ferromagnetic or short-range spin glass[21]/spin ice[22] transition. They consistently suggest no spin freezing at least down to 60 mK, despite the AF exchange energy ($\theta_w$ ~ -4 K). In addition, YbMgGaO$_4$ is a good insulator with a room temperature resistance greater than 20 MΩ. These results imply that the compound is a new QSL candidate.

The discrepancy between the experimental observations and numerical results[23] of the spin-1/2 triangular Heisenberg or XXZ model is attributed to the following possibilities. First, the effective-1/2 exchange matrix between neighbor Yb$^{3+}$ may significantly deviate from the ideal Heisenberg or XXZ model[24], and may cause much stronger spin frustration and fluctuation at low temperatures. And we would like to discuss this in detail in the further experimental and theoretical works basing on large single crystals. Second, slight bond randomness may exist in YbMgGaO$_4$ according to the crystal structure model II, which may influence the GS. The bond randomness can prevent spin ordering in THAF according to the exact diagonalization (ED) study[25]. Third, the next-nearest-neighbor and/or longer exchange interactions may stabilize a spin disordered GS against AF states[26,27]. Finally, the multiple-spin[28]/ring exchanges[29] can also prevent Néel long-range ordering in a triangular lattice. Quantum fluctuations play an important role in the calculated ordered GS because of the geometrical frustration even in Heisenberg case[5]. This fact means that some seemingly small perturbations may be critical in the determination of the GS of the frustrated spin system at low temperatures.

The heat capacity is a highly sensitive probe for the low-energy excitations from the GS. The availability of a perfect nonmagnetic reference compound LuMgGaO$_4$ enables an accurate exclusion of the lattice heat capacities without any fitting. Such advantage is absent in most of the existing QSL candidates. The measured heat capacities of LuMgGaO$_4$ well follow the Debye law with a Debye temperature ~ 151 K (Fig. 3a). Thus the exact magnetic heat capacities of YbMgGaO$_4$ can be precisely extracted by directly subtracting the lattice contributions, i.e., the heat capacities of LuMgGaO$_4$, from those of YbMgGaO$_4$ (Fig. 3b). The heat capacity measurements were performed up to 30 K, at which the spin-1/2 entropy has been fully released as $T \gg |\theta_w|$, and the Yb$^{3+}$ ions remain in the Kramers doublet GSs with effective spin-1/2 as $C_m$(~30 K, 0 T) ~ 0 (Fig. 3b). As the temperature decreases, the magnetic heat capacities of YbMgGaO$_4$ exhibit a broad hump, whose position is almost field-independent ($\mu_0H \leq 2$ T), and shifts to a higher temperature with further increasing applied magnetic fields ($\mu_0H \geq 4$ T). The broad hump centered at 2.4 K (under 0 T) may suggest a crossover into a QSL state[30]. At $T < 2$ K, the magnetic heat capacities well follow power-law temperature dependences down to the lowest measurement temperatures (Fig. 3b). The fitted power exponent, γ ~ 0.7, approaches the theoretical value of 2/3 reported in the THAF spin liquid with ring exchanges[29]. It is another evidence that YbMgGaO$_4$ is a new strongly correlated QSL candidate. γ increases up to 2.7 under 9 T (the inset of Fig.3c) possibly because of the gradually overcoming of the 2D quantum spin correlations. We have also tried to fit the low temperature heat capacities (60 to 83 mK, 0 T) with a gapped spectral function, C*exp[-ΔE/$T$]. The fitted energy gap, ΔE = 0.0469(1) K, is no more than ~ |θ$_w$|/100 (Fig. 3c). Moreover, the susceptibilities (Fig. 2d) show no downward trend to zero down to 0.48 K, which is much lower than the hump temperature. These consistently suggest that YbMgGaO$_4$ is a QSL candidate with an excitation gap no more than ~ |θ$_w$|/100.

The released magnetic entropies from the lowest measurement temperature to 30 K are exactly ~ Rln2/mol Yb$^{3+}$ in YbMgGaO$_4$ (Fig. 3d). In other words, the residual spin entropy at 60 mK is almost zero. This means that the gapless QSL candidate has a disordered but macroscopically non-degenerate GS at low temperatures, and thus the third law of thermodynamics remains unviolated. Moreover, the near-zero upper limit of residual entropy (~ 0.6 %) at 60 mK almost excludes the possibility of magnetic transitions at lower temperatures and indicates that a possible QSL GS is experimentally achieved in YbMgGaO$_4$, since there are no sufficient entropies for spin symmetry breaking any more below 60 mK.

The neighbor exchange energy, |θ$_w$|, is on the order of several Tesla. The magnetization of YbMgGaO$_4$ under a stable magnetic field at very low temperatures

is critical and interesting. Unconventional quantum spin states, such as quantum magnetization plateaus[8-10], can be induced by applied fields.

The careful magnetization measurements of YbMgGaO$_4$ at 0.5 K under 0 to 7 T, as well as at higher temperatures, are shown in Fig. 4a. The derived susceptibilities (dM/dH) show a clear and anomalous plateau under fields of 1.6 to 2.8 T at 0.5 K, and this plateau completely disappears at higher temperatures (1.9, 2.5 and 4.2 K), as shown in Fig. 4b. The unconventional magnetization behavior was confirmed by repeating the measurements several times. The magnetization curve at 0.5 K (Fig. 4a) is very similar to that of the triangular lattice compound C$_6$Eu along the *c*-axis at 4.2 K[31]. In the case of C$_6$Eu, this behavior has been attributed to traces of the 1/3 plateau[32]. It should be pointed out that the fitted power exponent γ also shows an unusual quasi plateau (γ ~ 1.5) in the similar magnetic-field range at extremely low temperatures (inset of Fig. 3c). The unusual χ-plateau must be related to a field-induced unconventional quantum state at extremely low temperatures ($\leq |\theta_w|/10$).

The one-third quantum magnetization plateau, as well as higher-fraction quantum states, have been reported in Néel ordering phases ($T < T_N$) in spin-1/2 triangular-lattice antiferromagnets such as Ba$_3$CoSb$_2$O$_9$ and Cs$_2$CuBr$_4$, and the Néel transitions are caused by the interlayer coupling[8-10]. To the best of our knowledge, YbMgGaO$_4$ is the first example that exhibits the abnormal magnetization behavior in a paramagnetic phase in a 2D lattice. To fully understand the abnormal magnetization behavior, further theoretical and experimental studies are required.

A new triangular antiferromagnet with effective spin-1/2 and perfect $R\bar{3}m$

symmetry, YbMgGaO$_4$, was successfully synthesized. The new compound overcomes some critical disadvantages in the existing materials, such as spatially anisotropic intralayer exchange interactions, magnetic defects, interlayer exchange coupling, and antisymmetric DM interactions. Despite the significant AF exchange coupling between neighbor spins ($\theta_w \sim$ -4 K), no spin freezing was observed at least down to 60 mK, suggesting that YbMgGaO$_4$ is a new QSL candidate. The gapless low-energy excitations from the GS are evidenced by the low-$T$ power-law temperature dependence of heat capacity and nonzero susceptibility. Almost zero residual spin entropy suggests that the new compound extremely approaches a disordered and possible gapless QSL GS at low temperatures. An unconventional field-induced quantum state featuring an abnormal susceptibility plateau was observed for the first time, where the frustrated spin system was still paramagnetic at 0.5 K.

**Methods**

Yb$_x$Lu$_{1-x}$MgGaO$_4$ ($x$ = 1, 0.4, 0.16, 0.08, 0.04 and 0) white powder samples were synthesized from stoichiometric mixtures of Yb$_2$O$_3$ (99.99%), Lu$_2$O$_3$ (99.9%), MgO (99.99%) and Ga$_2$O$_3$ (99.999%). The mixtures were heated in air to 1450 °C for 4 days, with an intermediate grinding. The phase purities of all of the samples were confirmed by powder X-ray diffraction (XRD) (Bruker D8 Advance, 40 kV, 40 mA) before further study. High-intensity and monochromatic synchrotron XRD was used to more precisely determine the crystal structures of YbMgGaO$_4$ and LuMgGaO$_4$ at the diffraction station (4B9A) of the Beijing Synchrotron Radiation Facility (BSRF).

The General Structure Analysis System (GSAS) program was used for Rietveld crystal structure refinements[33]. No structural transitions were observed from the XRD patterns (Rigaku, 40 kV, 200 mA) at least to 12 K in YbMgGaO$_4$. Magnetization measurements at 0.48 to 300 K under 0 to 14 T were performed using a superconducting quantum interference device (SQUID) magnetometer (Quantum Design Magnetic Property Measurement System, MPMS) and a vibrating sample magnetometer (VSM) (Quantum Design Physical Property Measurement System, PPMS). The magnetizations were measured at 0.48 to 2 K using a He3 refrigeration system. The magnetizations of Yb$_x$Lu$_{1-x}$MgGaO$_4$ ($x$ = 1, 0.4, 0.16, 0.08 and 0.04) were carefully corrected for the weak contribution (< 3 %) of the brass sample holder to the diamagnetic background. Heat capacity measurements at 0.06 to 30 K under 0 to 9 T were performed using a Quantum Design PPMS. YbMgGaO$_4$/ LuMgGaO$_4$ powder sample mixed with Ag (99.9%) powder (mole ratio = 1:3) was dye-pressed into hard disks to facilitate thermal equilibration. The heat capacities of YbMgGaO$_4$ and LuMgGaO$_4$ were carefully corrected for the separately measured Ag and other addenda (including heat-conducting glue) contributions. The heat capacity measurements between 3 K and 60 mK were performed using a dilution refrigeration system.

**Acknowledgements**

We thank Gang Chen, Rong Yu and Hechang Lei for helpful discussions. This work was supported by the NSF of China and the Ministry of Science and Technology of China (973 projects: 2011CBA00112 and 2012CB921701). Q.M.Z. and Y.S.L were supported by the Fundamental Research Funds for the Central Universities, and the Research Funds of Renmin University of China.

**Figure legends**

**Figure 1** (a) Polyhedral structure of YbMgGaO$_4$. The black dashed lines indicate the unit cell. (b) Top view of the triangular layer of YbO$_6$ octahedra. (c) Local crystal structure around the Yb$^{3+}$ Kramers ions.

**Figure 2** (a) Temperature dependence of magnetization under zero-field-cooling (ZFC) and 1 T in Yb$_x$Lu$_{1-x}$MgGaO$_4$. Inset: zoomed view of the low-temperature data. (b) Magnetic field dependence of magnetization at 2.5 K. The colored dash lines show Van Vleck paramagnetism extracted from the linear-field-dependent magnetization data (> 10 T). (c) Curie-Weiss fits of magnetization data at low temperatures (< 20 K). Inset: fitted (AF) Weiss temperatures. (d) Susceptibilities measured under ZFC and FC from 0.48 to 30 K. No splitting between the ZFC and the FC data was observed at temperatures above 0.48 K. Both the cooling field H$_c$ and the measurement field H$_m$ are 100 Oe. Inset: complete magnetic loop measured at 0.5 K. In both the first and third quadrants, the data collected under increasing fields are perfectly overlapped by those collected during decreasing field.

**Figure 3** (a) Temperature dependences of total heat capacity under different magnetic fields in YbMgGaO$_4$ and LuMgGaO$_4$. The dashed curve denotes the Debye heat capacity. (b) Temperature dependences of magnetic heat capacity under different magnetic fields in YbMgGaO$_4$. The colored dash lines show the power law fits to the

low-*T* magnetic heat capacities. (c) Magnetic heat capacity vs. $1/T$ in YbMgGaO$_4$. The red dash line shows the gapped spectral function fit to the low-*T* magnetic heat capacities under 0 T. Inset: fitted power exponent $\gamma$. (d) Temperature dependences of integral magnetic entropy under different magnetic fields in YbMgGaO$_4$.

**Figure 4** (a) Magnetic field dependence of magnetization at 0.5, 1.9, 2.5 and 4.2 K. The dashed line represents Van Vleck magnetism. (b) Magnetic field dependence of susceptibilities (dM/dH) at 0.5, 1.9, 2.5 and 4.2 K. The dashed lines show the interval where almost constant susceptibilities are observed.

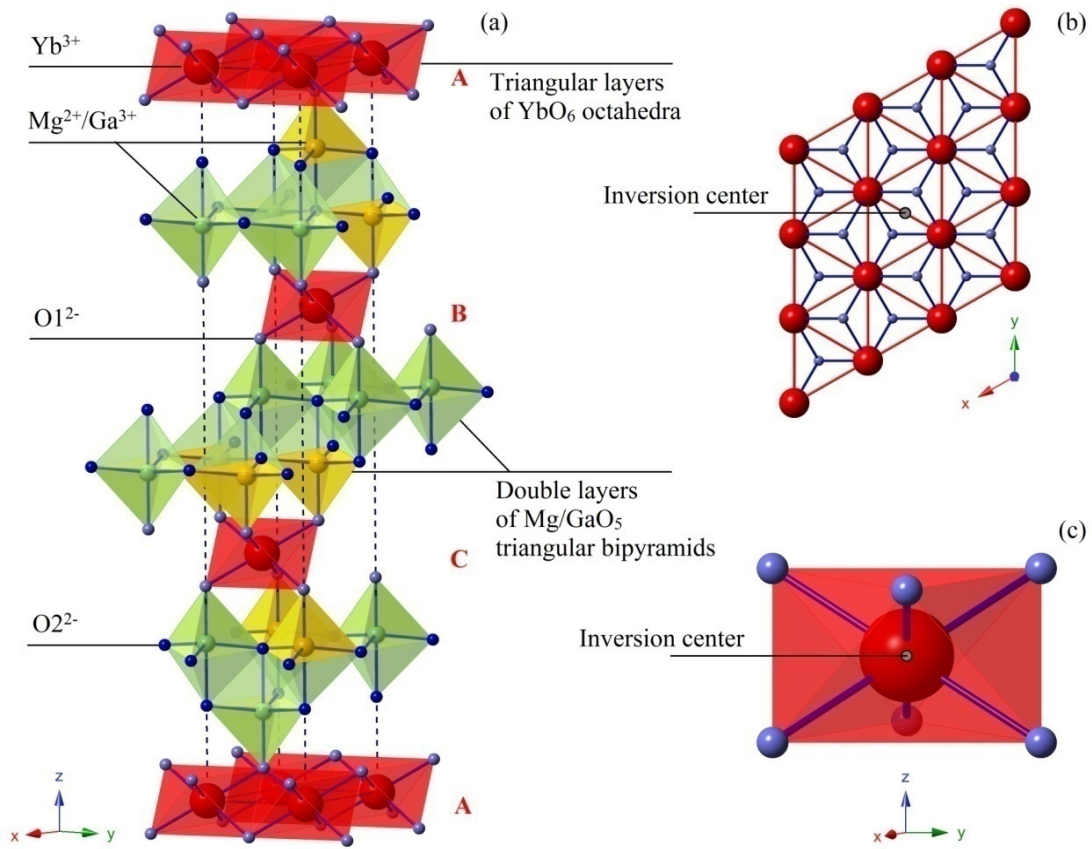

**Figure 1**

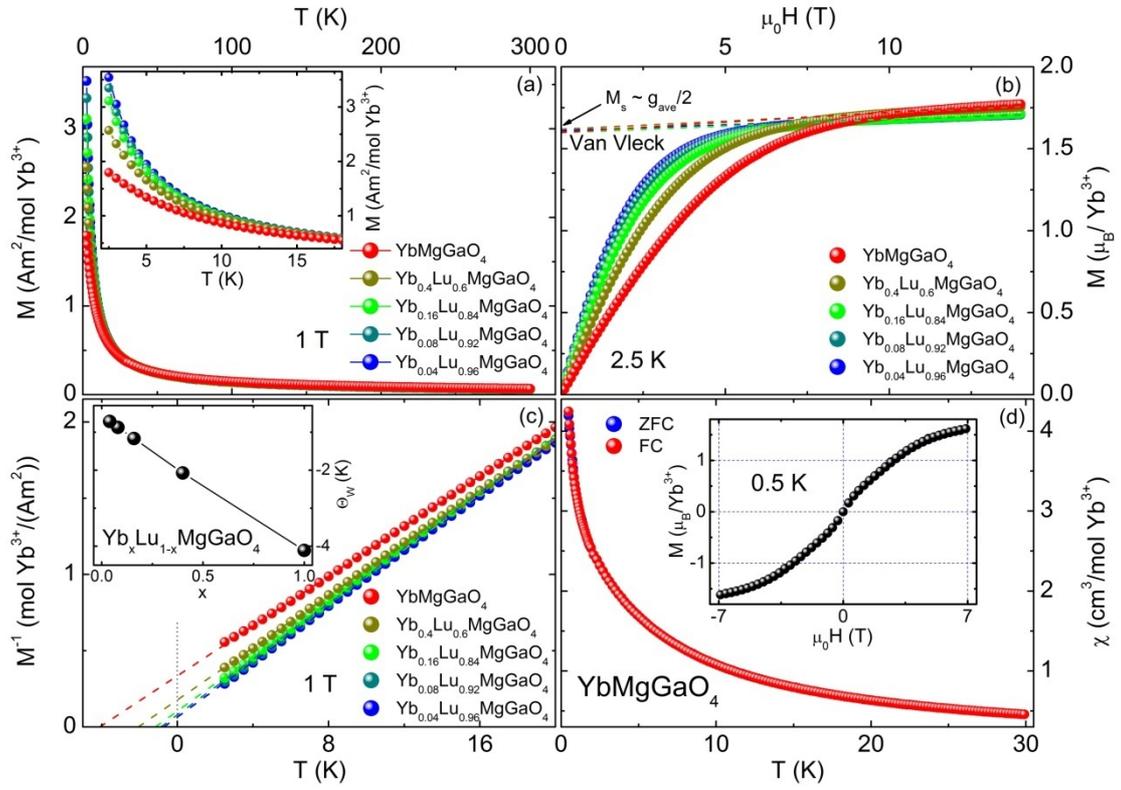

**Figure 2**

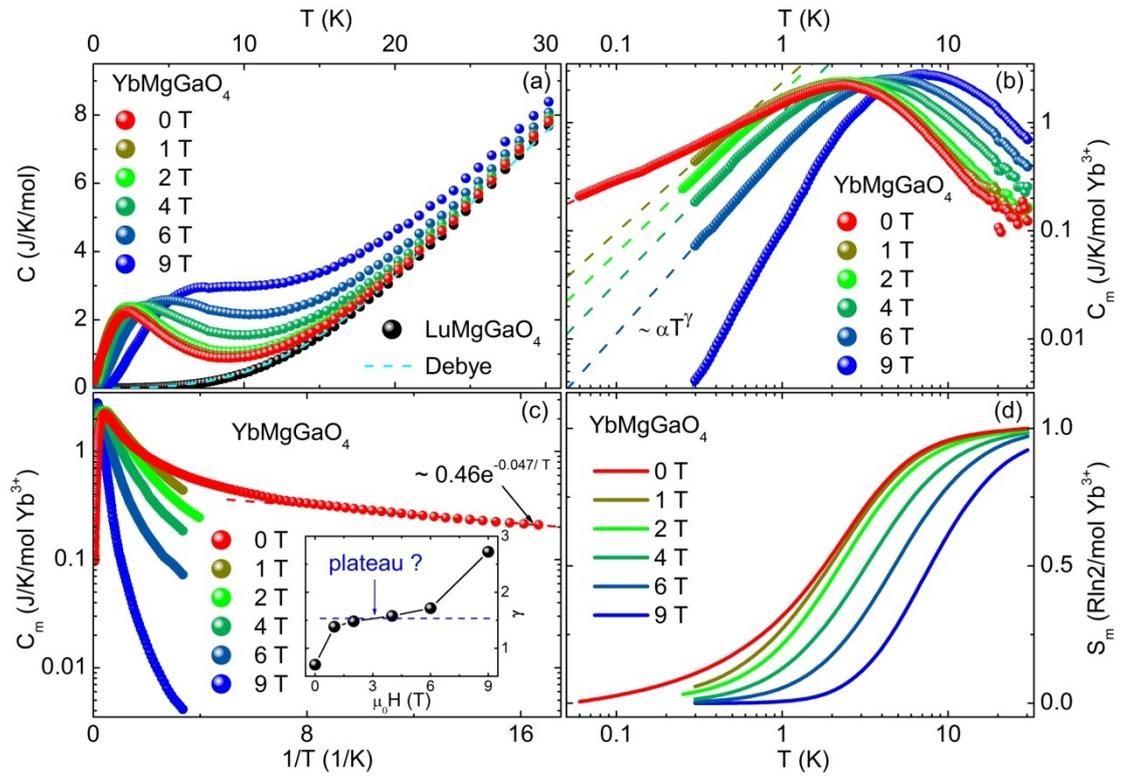

**Figure 3**

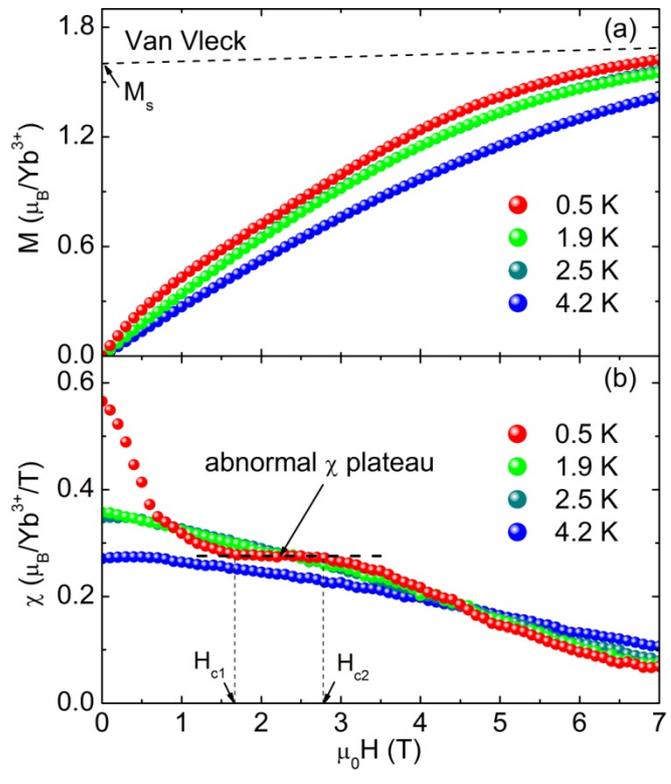

**Figure 4**

# Supplemental Material

# Gapless quantum spin liquid ground state in the two-dimensional spin-1/2 triangular antiferromagnet YbMgGaO$_4$

Yuesheng Li[1], Haijun Liao[2], Zhen Zhang[3], Shiyan Li[3], Feng Jin[1], Langsheng Ling[4], Lei Zhang[4], Youming Zou[4], Li Pi[4], Zhaorong Yang[5], Junfeng Wang[6], Zhonghua Wu[7], Gang Chen[8] and Qingming Zhang[*,1, 9]

[1]Department of Physics, Renmin University of China, Beijing 100872, P. R. China
[2]Institute of Physics, Chinese Academy of Sciences, Beijing100190, P. R. China
[3]State Key Laboratory of Surface Physics, Department of Physics, and Laboratory of Advanced Materials, Fudan University, Shanghai 200433, P. R. China
[4]High Magnetic Field Laboratory, Chinese Academy of Sciences, Hefei 230031, P. R. China
[5]Institute of Solid State Physics, Chinese Academy of Sciences, Hefei 230031, P. R. China
[6]Wuhan National High Magnetic Field Center, Wuhan 430074, P. R. China
[7]Institute of High Energy Physics, Chinese Academy of Science, Beijing 100049, P. R. China
[8]
[9]Department of Physics and Astronomy, Collaborative Innovation Center of Advanced Microstructures, Shanghai Jiao Tong University, Shanghai 200240, P. R. China
[*]e-mail: qmzhang@ruc.edu.cn

**We present here:**

1. Material synthesis and measurement details
2. Synchrotron X-ray diffractions and structure refinements of YbMgGaO$_4$ and LuMgGaO$_4$
3. Low-temperature XRD patterns of YbMgGaO$_4$.
4. ESR spectra of YbMgGaO$_4$ and Yb$_{0.04}$Lu$_{0.96}$MgGaO$_4$ samples

## 1. Material synthesis and measurement details

Yb$_x$Lu$_{1-x}$MgGaO$_4$ ($x$ = 1, 0.4, 0.16, 0.08, 0.04 and 0) powder samples were prepared via the following chemical reaction:

$$x\text{Yb}_2\text{O}_3 + (1-x)\text{Lu}_2\text{O}_3 + 2\text{MgO} + \text{Ga}_2\text{O}_3 = 2\text{Yb}_x\text{Lu}_{1-x}\text{MgGaO}_4$$

Stoichiometric mixtures of high-purity Yb$_2$O$_3$ (99.99%), Lu$_2$O$_3$ (99.9%), MgO (99.99%) and Ga$_2$O$_3$ (99.999%) were ground, pressed into tablets and heated in air at 1450 °C for 4 days with an intermediate grinding step. Almost no sample mass loss was observed after heating.

To remove the influence of a possible preferred orientation of the powder samples during the crystal-structure refinements, the solid-phase, synthesized samples were carefully ground into powders with particle sizes smaller than ~ 5 μm, as determined by microscope observations. The powder samples were diluted with a moderate amount of glue, packed into capillaries (D = 0.2 mm) and prepared for X-ray diffraction using the MYTHEN detector at the diffraction station (4B9A) of the Beijing Synchrotron Radiation Facility.

For the magnetization and electron spin resonance (ESR) spectrum measurements, at least two independent powder samples were prepared and measured for each magnetic Yb$_x$Lu$_{1-x}$MgGaO$_4$ composition. In addition, no clear effects caused by the possible preferred orientation of the samples were observed from the measurements.

The ESR measurements were performed using a Bruker EMX plus 10/12 CW-spectrometer at X-band frequencies (f ~ 9.4 GHz); the spectrometer was equipped with a continuous He gas-flow cryostat.

The residuals in all fits and refinements are defined as

$$R_p = \frac{\sum |y_o - y_c|}{\sum |y_o|} \quad (1)$$

where, $y_o$ and $y_c$ are observed and calculated values respectively. International system of units (SI) was used.

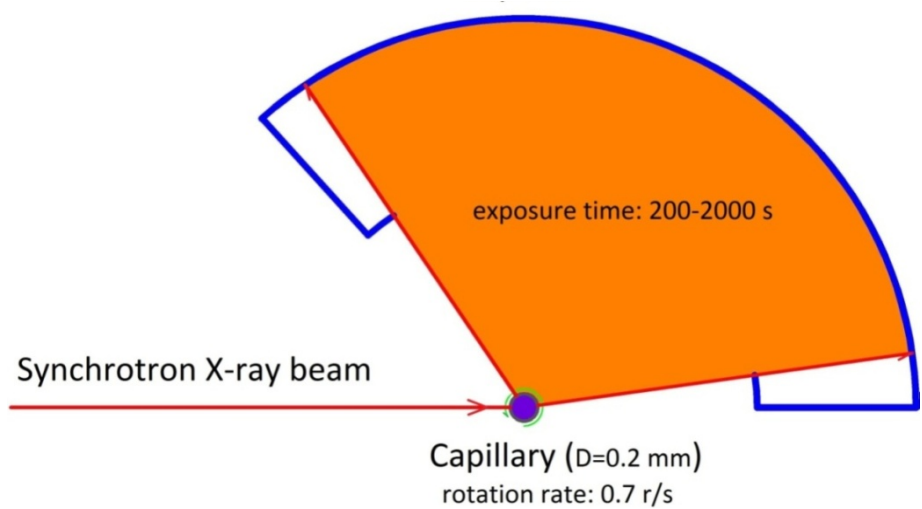

**Figure S1.** Diagram of the MYTHEN detector installed at the diffraction station (4B9A) of the Beijing Synchrotron Radiation Facility.

2. **Synchrotron X-ray diffractions and structure refinements of YbMgGaO$_4$ and LuMgGaO$_4$**

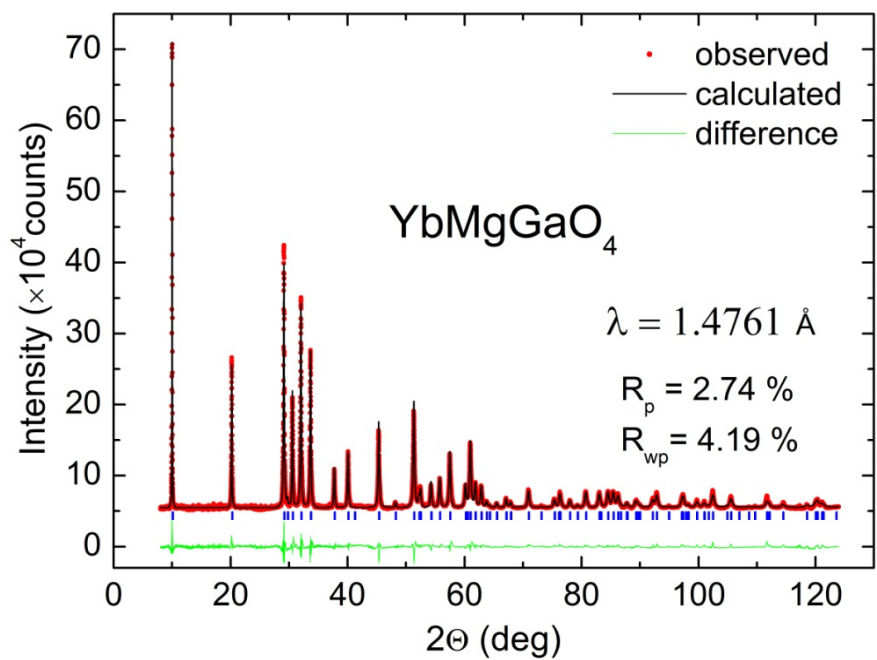

**Figure S2.** Synchrotron X-ray diffraction and final Rietveld refinement profiles for YbMgGaO$_4$ at 300 K.

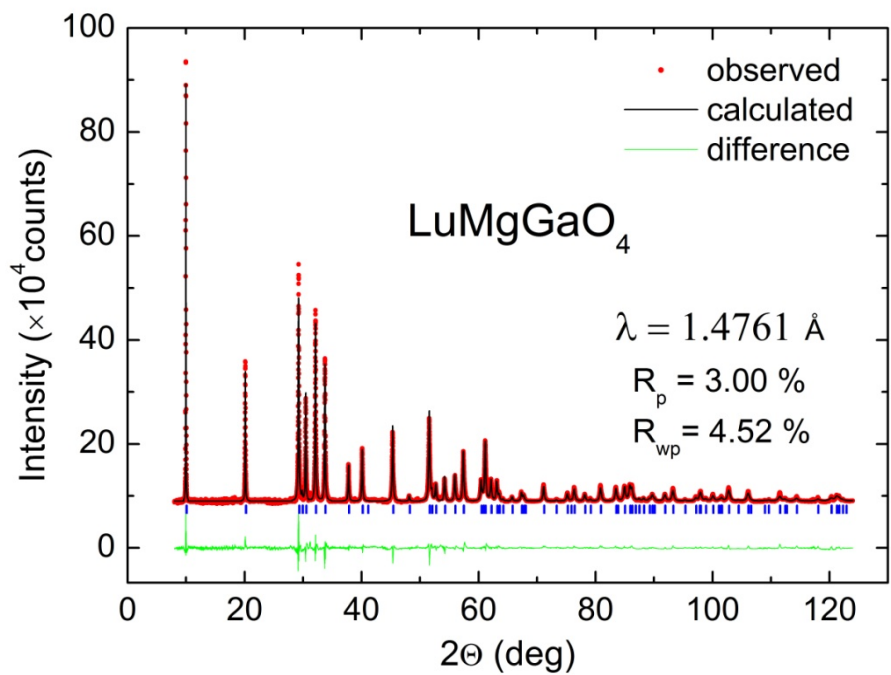

**Figure S3.** Synchrotron X-ray diffraction and final Rietveld refinement profiles for LuMgGaO$_4$ at 300 K.

**Table S1.** Refined crystal structures of YbMgGaO$_4$ and LuMgGaO$_4$

| Compound | | YbMgGaO$_4$ | | LuMgGaO$_4$ | |
|---|---|---|---|---|---|
| Model | | I | II | I | II |
| Space group | | $R\bar{3}m$ | | | |
| Lattice | $a$ (Å) | 3.40212(8) | 3.40282(6) | 3.38750(5) | 3.38750(5) |
| | $c$ (Å) | 25.1191(6) | 25.1243(5) | 25.2069(4) | 25.2071(4) |
| Yb/Lu$^{3+}$ | Fraction | 1 | 0.5 | 1 | 0.5 |
| | $z$ | 0 | 0.00439(6) | 0 | 0.00425(6) |
| | Uiso (×100) | 1.016(18) | 0.751(21) | 0.981(18) | 0.637(20) |
| Mg$^{2+}$/Ga$^{3+}$ | Fraction | 0.5 | 0.5 | 0.5 | 0.5 |
| | $z$ | 0.214902(29) | 0.214918(29) | 0.215397(30) | 0.215385(30) |
| | Uiso (×100) | 0.216(26) | 0.301(26) | 0.332(27) | 0.226(27) |
| O1$^{-2}$ | $z$ | 0.29260(10) | 0.29221(10) | 0.29177(11) | 0.29213(11) |
| | Uiso (×100) | 0.92(8) | 0.72(8) | 1.15(8) | 1.04(8) |
| O2$^{-2}$ | $z$ | 0.12967(9) | 0.12993(9) | 0.12967(10) | 0.12971(10) |
| | Uiso (×100) | 0.54(8) | 0.47(7) | 1.02(8) | 0.68(7) |
| Residuals | $R_{wp}$ | 0.0419 | 0.0412 | 0.0452 | 0.0442 |
| | $R_p$ | 0.0274 | 0.0268 | 0.0300 | 0.0292 |

3. **Low-temperature XRD patterns of YbMgGaO$_4$.**

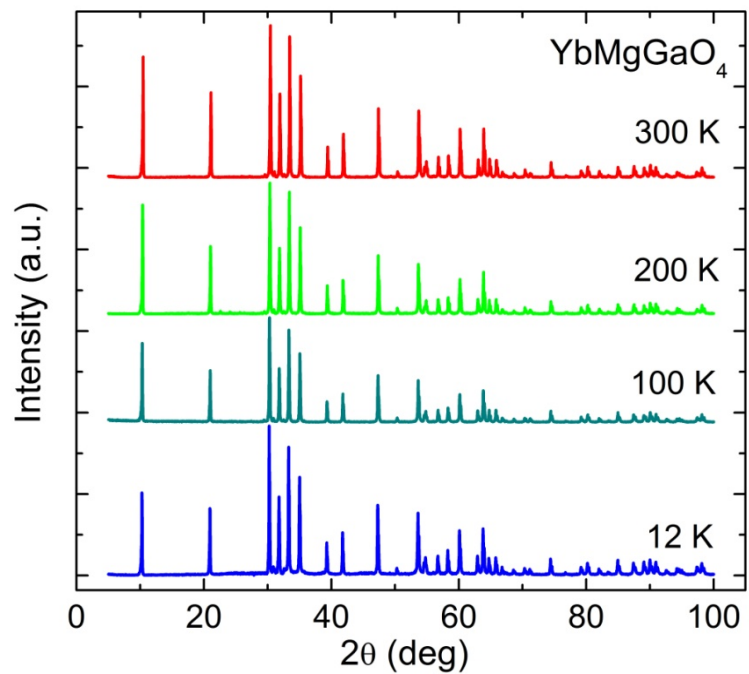

**Figure S4.** X-ray (Cu-K$_\alpha$) diffraction profiles for YbMgGaO$_4$ at 300, 200, 100 and 12 K respectively. No additional reflections are observed at temperatures down to 12 K, indicating that no structural transitions occurred.

## 4. ESR spectra of YbMgGaO$_4$ and Yb$_{0.04}$Lu$_{0.96}$MgGaO$_4$ samples

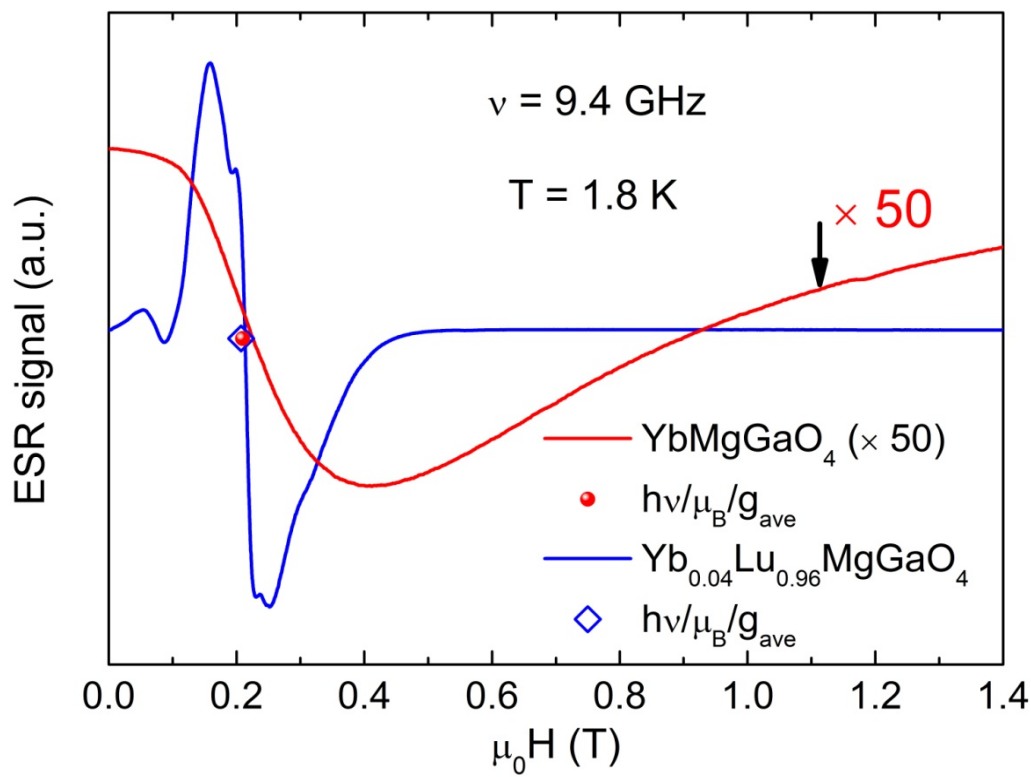

**Figure S5.** X-band (9.4 GHz) Yb$^{3+}$ first-derivative absorption ESR spectra for YbMgGaO$_4$ and Yb$_{0.04}$Lu$_{0.96}$MgGaO$_4$. The narrow and intense hyperfine lines[1] observed in the quasi-free spin compound Yb$_{0.04}$Lu$_{0.96}$MgGaO$_4$ completely disappear in the spectrum of YbMgGaO$_4$, suggesting that no observable isolated Yb$^{3+}$ or magnetic defects are present in YbMgGaO$_4$.